\documentclass{webofc}
\usepackage[varg]{txfonts}   
\usepackage{gensymb}

\begin{document}
\title{\textit{I see you, do you see me?} Perception-based crowdedness and behavioral responses in pedestrian dynamics}

\author{
\firstname{Igor} \lastname{Hołowacz}\inst{1}\fnsep
\and
\firstname{Pratik} \lastname{Mullick}\inst{2}\fnsep
\thanks{\email{pratik.mullick@pwr.edu.pl}}
}

\institute{
Department of Artificial Intelligence, Wrocław University of Science and Technology, Poland
\and
Department of Operations Research and Business Intelligence, Wrocław University of Science and Technology, Poland
}

\abstract{%
  \textit{Problem definition}: Pedestrian traffic is commonly characterized using local density, yet the interactions experienced by individuals also depend on the relative positions and perceptual relevance of surrounding pedestrians. This raises the question of whether behavioral relationships inferred from local crowdedness are robust to the way perceptual anisotropy is represented, and how interaction geometry shapes pedestrian adaptation over time. \textit{Methodology}: We analyze experimental pedestrian crossing flows over crossing angles from 0 to 180 degrees using a distance-weighted measure of local crowdedness. Perceptual anisotropy is varied by progressively reducing the contribution of pedestrians outside the focal pedestrian's field of view. We study the temporal evolution of crowdedness and its relationships with velocity, directional deviation, and acceleration. \textit{Results}: We find that perceptual anisotropy primarily changes the numerical scale of crowdedness, while the qualitative dynamics, temporal progression, and crossing-angle dependence remain largely preserved. Pedestrians deviate appreciably from their expected group directions, but changes between successive walking directions remain small, indicating adaptation through smooth, incremental corrections rather than abrupt turns. Acceleration dynamics further show an asymmetry between disruption and recovery: initial deceleration varies strongly with crossing geometry, whereas subsequent recovery accelerations are considerably more similar across angles. The non-retracing trajectories in the behavioral phase spaces further show that similar instantaneous conditions can correspond to different stages of the interaction, revealing a dependence on the phase of interaction. \textit{Implications}: The robustness of the observed dynamics indicates that interaction geometry has a stronger influence on the organization of the crossing flows studied here than the precise perceptual weighting used to quantify local crowdedness. More broadly, the results show that dynamic fundamental diagrams provide a more complete characterization of transient pedestrian interactions than conventional relationships based on instantaneous state variables alone.
}
\maketitle
\section{Introduction}

\textit{I saw you long before our paths crossed. Across the road, among the noise and motion of the street, your steps had already begun to determine mine. I slowed, shifted, made space, convinced that, at the proper moment, you too would see me. The moment came. You did not. You walked past as if I had never been there, and our paths, which I had watched drawing closer for so long, separated without ever meeting. Only then did I understand that I had been avoiding you from the beginning.}

Such ordinary moments, repeated countless times in streets, stations, corridors, and public squares, shape the way pedestrian traffic moves \cite{helbing2001traffic,schreckenberg2002pedestrian}. Walking through a crowd is more than simply reaching a destination; it is a continuous process of noticing, anticipating, and adjusting to the movements of other agents or obstacles. Shaped by how pedestrians perceive and interpret the movements of others \cite{gibson1958visually,batty1997predicting,turner2002encoding,gulikers2013effect,wirth2023neighborhood}, a brief glance, a subtle change in speed, or a small shift in direction can prevent collision and maintain the rhythm of collective motion \cite{schrater2000mechanisms,hopkins2004braking,huber2014adjustments}. Yet when many such local adjustments occur simultaneously, we see self-organizing patterns \cite{helbing2012social,moussaid2012traffic,gu2025emergence}, including lanes \cite{helbing2005self,feliciani2016empirical,murakami2021mutual,khelfa2022heterogeneity,bacik2023lane}, stripes \cite{mullick2022analysis,zanlungo2023amacroscopic,zanlungo2023bmacroscopic}, and other forms of coordinated structures \cite{hoogendoorn2005pedestrian,seyfried2009new,zhang2013experimental}. The dynamics of pedestrian flows are therefore shaped not only by the physical space occupied by individuals, but also by the behavioral processes through which they respond to and negotiate interactions with others \cite{moussaid2011simple,murakami2021mutual,dachner2022visual,corbetta2023physics}. Understanding how these individual adaptations interact with the geometry of pedestrian encounters is essential for explaining how collective organization emerges and how pedestrian traffic evolves under different flow configurations.

Pedestrian dynamics have also been studied extensively through simulations \cite{dogbe2012modelling,duives2013state,van2021algorithms,korbmacher2023time}, which provide a controlled framework for investigating how individual interaction rules give rise to collective flow patterns under a wide range of conditions. A variety of simulation frameworks have been developed to generate human trajectories, such as the social force model \cite{helbing1995social}, cellular automata and floor-field models \cite{burstedde2001simulation}, alignment-based models such as the Vicsek model \cite{vicsek1995novel}, velocity-based collision-avoidance approaches \cite{kim2015velocity,xu2021anticipation}, and cognitive models based on visual perception and anticipation \citep{ondvrej2010synthetic,moussaid2011simple,HOOGENDOORN2014684,dachner2022visual}. More recently, local density has also been incorporated into heuristic pedestrian models \cite{koh2011modeling,dietrich2014gradient,xiao2016pedestrian}, allowing agents to adapt their walking speed to the evolving state of their immediate surroundings. A major motivation for such studies has been to understand the mechanisms underlying dangerous crowd phenomena and to develop strategies for preventing crowd disasters \cite{helbing2002crowd,helbing2007dynamics,Helbing2012love,HAGHANI2018253,feliciani2022introduction,feliciani2023trends,gu2025emergence}.

%At the collective level, pedestrian traffic has traditionally been characterized through macroscopic quantities such as density and velocity, and particularly through the relationship between them. The resulting fundamental diagrams provide a central framework for describing how pedestrian motion changes as local crowding increases, and have been widely used to compare different flow configurations and levels of congestion. However, such descriptions depend critically on how the local state surrounding a pedestrian is quantified, raising the question of whether conventional, purely geometric measures of density fully capture the interactions experienced by individuals within a moving crowd.

%pedestrian crowd dynamics (very generic paragraph, with focus on safety-security analysis)

Beyond its role in pedestrian simulation, density is one of the principal quantities used to analyze pedestrian traffic \cite{STEFFEN20101902,schauer2014estimating,duives2013state,Wirz2013,DUIVES2015162,tordeux2015,rao2015estimation,Nagao2018,ding2020crowd,WANG2023104400,MULLICK2025130251}. Its relationship with walking speed, commonly represented through fundamental diagrams \cite{seyfried2005,GEROLIMINIS2008759,KEYVANEKBATANI20121393,PAETZKE2022,mullick2025classifying}, provides a standard description of how pedestrian motion changes under increasingly crowded conditions. However, such descriptions depend critically on how the local state surrounding a pedestrian is quantified, raising the question of whether conventional, purely geometric measures of density could solely capture the interactions experienced by individuals within a moving crowd. On the other hand, microscopic metrics such as the avoidance and intrusion numbers \cite{cordes2024classification,cordes2024dimensionless,mullick2025classifying} capture individual-level processes related to collision anticipation and personal-space preservation.

A further consideration is that the local interaction state may depend not only on the physical configuration of surrounding pedestrians, but also on how they are perceived. Visual information used to control motion is shaped by factors such as relative position, interpersonal distance, and occlusion \cite{gibson1958visually,turner2002encoding}. In particular, pedestrians ahead and behind are unlikely to be equally relevant to navigation \cite{helbing2012social}, motivating anisotropic descriptions of the local interaction state \cite{HOOGENDOORN2014684,moussaid2011simple,gulikers2013effect,corbetta2023physics,GARCIA2023128461}.

It is important, however, to distinguish between anisotropy incorporated into the interaction mechanism itself and anisotropy introduced into the observational measure used to characterize local crowdedness. In the latter context, restricting crowdedness measures to a pedestrian's field of view has been shown to substantially alter the resulting behavioral relationships in unidirectional bottlenecks \cite{DUIVES2015162}. By contrast, a study of pedestrian zebra crossings found that varying the parametrization of anisotropy changed the numerical interaction measure without necessarily altering the qualitative behavioral structure \cite{WANG2023104400}. These contrasting observations raise a broader question of how robust the inferred relationships between local crowdedness and pedestrian behavior are to the particular representation of perceptual anisotropy.

In this study, we address this question by investigating how the representation of perceptual anisotropy influences the inferred relationship between local crowdedness and pedestrian behavior. Rather than treating individuals outside the field of view as either fully relevant or entirely irrelevant, we progressively vary their contribution to a distance-weighted measure of crowdedness and test whether the resulting behavioral relationships remain qualitatively robust. We further study how changes in the local crowdedness are associated with pedestrian velocity, directional adaptation, and acceleration across different stages of the interaction.

Pedestrian crossing-flows provide a particularly suitable setting for this investigation, as the geometry of interaction can be systematically varied while the resulting behavioral adaptation and collective organization are observed. We analyze experimental data from bidirectional crossing flows \cite{pedinteract,mullick2022analysis} with crossing angles ranging from $0\degree$ to $180\degree$, comprising of a total of 116 trials. The experimental trials were conducted with two groups of participants crossing through each other and their trajectories were recorded using motion capture at 120 Hz. Previous analyses of these data have revealed crossing-angle dependence of velocity-density relationship \cite{mullick2025classifying}, as well as the emergence of self-organized stripe patterns \cite{mullick2022analysis}. The dataset therefore provides a very useful framework for investigating whether such geometry-dependent dynamics remain robust to the representation of perceptual anisotropy and how pedestrians adjust their speed and direction over the course of the interaction.

The remainder of the paper is organized as follows. Section \ref{sec:crowdedness} introduces the perception-based crowdedness measure and examines its temporal evolution and relationship with pedestrian velocity under different levels of perceptual anisotropy. Section \ref{sec:deviation} analyzes directional adaptations during crossing interactions using two complementary measures of pedestrian deviation, while Section \ref{sec:accln} investigates acceleration dynamics and their relationship with crowdedness. Section \ref{sec:implications} discusses the implications of these behavioral observations for the assessment of interacting pedestrian flows, and Section \ref{sec:conclusions} summarizes the main conclusions.

\section{Perceived crowdedness and velocity dynamics}\label{sec:crowdedness}

Pedestrian density is commonly defined as the number of individuals per unit area. While such definitions are useful for describing macroscopic crowd conditions, they may not fully describe how crowdedness is locally experienced by individual pedestrians during motion. In particular, pedestrian interactions are strongly influenced not only by the number of nearby individuals, but also by their relative distances and spatial arrangement. We here define \textit{perceived crowdedness} $\rho$, as the local level of crowding experienced by each pedestrian based on distance-weighted interactions with surrounding agents.

Moreover, pedestrian motion is understood to be perception-driven. Individuals primarily respond to the pedestrians they can see, anticipate, and interpret as potential obstacles, rather than to all surrounding pedestrians equally. Consequently, two environments with similar local crowdedness may produce very different movement behaviors depending on visibility and interaction geometry. Motivated by this, we also define perception-based variants of the index of crowdedness, where interactions are weighted according to the pedestrian’s field of view (FOV). This idea is in fact inspired from the original social force model \citep{helbing1995social} where an agent feels repulsive forces from other agents within their FOV and for agents outside their FOV, the forces were scaled down by a factor.

\begin{figure}[h!]
    \centering
    \includegraphics[width=0.6\linewidth]{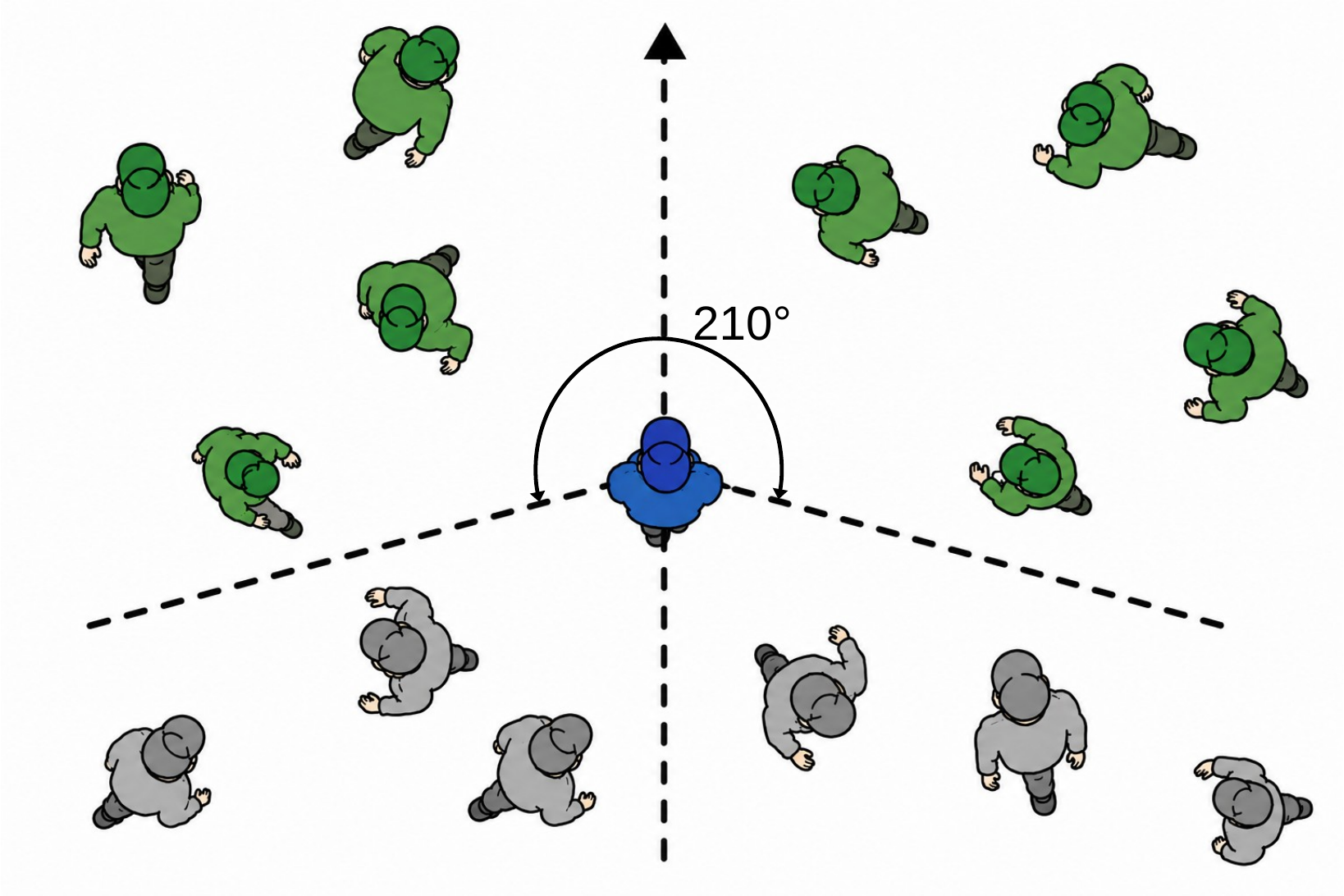}
    \caption{Schematic illustration of the field of view of a focal pedestrian. The blue pedestrian represents the focal agent, with the arrow indicating its direction of motion, while the dashed lines delimit its $210\degree$ field of view shown by the arc. Pedestrians within the field of view (green) are assigned a weight $w_{ij}=1$, whereas those outside it (gray) are assigned a weight $w_{ij}=c$, where $c\in[0,1]$ controls the level of perceptual anisotropy.}
    \label{fig:aniso}
\end{figure}

In our case, the perceived crowdedness $\rho^i$ of an agent $i$ (hereafter referred to as the focal agent) at time $t$ is defined as \begin{equation}
    \rho^i(t)
    = \sum_{j \neq i}
      \frac{w_{ij}}
           {\lVert \mathbf{r}_i(t) - \mathbf{r}_j(t) \rVert^2},
\end{equation} where $\mathbf{r}_k(t)$ denotes the position of agent $k$ at time $t$. The summation is performed over all other agents $j$. The weight factor $w_{ij}$ depends on whether $j$ is within the field of view of $i$ or not. To determine the field of view of agent $i$, we first estimate its direction of motion $\hat{\mathbf{d}}_i(t)$ as \begin{equation}
    \hat{\mathbf{d}}_i(t) =
    \frac{\mathbf{r}_i(t)-\mathbf{r}_i(t-1)}
         {\lVert\mathbf{r}_i(t)-\mathbf{r}_i(t-1)\rVert}.
\end{equation} Agent $j$ is said to be within the field of view of $i$ if the following condition is satisfied:
\begin{equation}
    \hat{\mathbf{d}}_i \cdot
    \frac{\mathbf{r}_j(t)-\mathbf{r}_i(t)}
         {\lVert\mathbf{r}_j(t)-\mathbf{r}_i(t)\rVert}
    \;\geq\; \cos\!\left(\frac{\phi}{2}\right).
\end{equation} The total field of view angle is taken as $\phi=210\degree$ \citep{traquair1927introduction, strasburger2020seven}. The weight $w_{ij}$ is then defined as \begin{equation*}
        w_{ij} =
    \begin{cases}
        1 & j \text{ inside FOV of } i \\
        c & j \text{ outside FOV of }i
    \end{cases}
\end{equation*} The parameter $c\in[0,1]$ controls the contribution of agents outside the field of view to the perceived crowdedness of focal agent. The cases $c=0$ and $1$ correspond to fully anisotropic (focal agent does not perceive agents outside its FOV) and isotropic (focal agent sees everyone) perception respectively, which was also studied in \cite{DUIVES2015162}. For $0<c<1$, agents outside the FOV contribute partially to the perceived crowdedness. The situation is shown schematically in Figure \ref{fig:aniso}. In this paper, we used $c=1$, $0.5$ and $0$, and the corresponding crowdedness measures are denoted by $\rho_c$. The goal is to investigate whether different levels of perceptual anisotropy influence pedestrian behavioral responses.

The analysis presented in the paper was restricted to the part of each trial associated with the crossing interaction. For each trial, $T_i$ and $T_f$ denote the approximate beginning and end of the interaction between the two pedestrian groups, respectively, and were identified as the times of the first and last edge cuts obtained from the edge-cutting algorithm \cite{mullick2022analysis}. To include the periods immediately before and after the interaction, we considered the trajectories from $T_i-2$ s to $T_f+2$ s. For the $\alpha=0\degree$ configuration, where no crossing interaction occurs and $T_i$ and $T_f$ are therefore not defined, the first and last four seconds of each trajectory were excluded to reduce boundary effects.

Since the duration of the interaction varies across trials, time was rescaled separately for each trial as $$t^\prime=\frac{t-T_i}{T_f-T_i},$$ such that $t^\prime=0$ and $t^\prime=1$ correspond to the approximate beginning and end of the interaction, respectively. The rescaled trajectories were divided into bins of width 0.01, and the quantities of interest were grouped according to their corresponding $t^\prime$ values. Within each bin, values were averaged across pedestrians and trials for each crossing angle $\alpha$. This procedure provides a common temporal scale for comparing the evolution of crowdedness, velocity, directional deviation, acceleration, and their mutual relationships across trials with different durations.

Our results show a common and consistent picture regarding the role of perceptual anisotropy in crossing pedestrian flows. The qualitative behavior remains remarkably robust to changes in the anisotropy parameter $c$. The principal effect of decreasing $c$ is a systematic reduction in the estimated crowdedness, which corresponds to the reduced contribution of pedestrians outside the field of view. However, the temporal trends, the relative ordering of crossing angles, and the overall dynamical structure remain largely unchanged. These observations suggest that perceptual anisotropy acts primarily as a rescaling of the crowdedness measure rather than altering the underlying collective dynamics, as also found in a previous empirical work \cite{WANG2023104400}.

\begin{figure*}
    \centering
    \includegraphics[width=\linewidth]{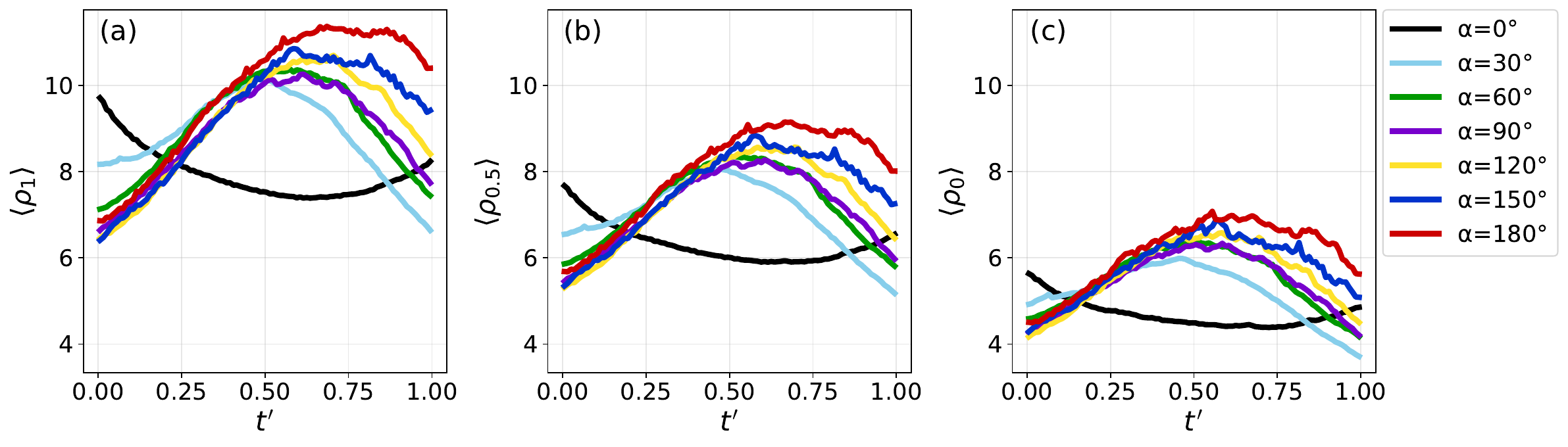}
    \caption{Variation of average values of indices of crowdedness (a) $\rho_1$, (b) $\rho_{0.5}$ and (c) $\rho_0$ as functions of the scaled time $t'$ for different values of the crossing angle. The averages are estimated over all the agents and all the experimental trials for a particular crossing angle.}
    \label{fig:fig1}
\end{figure*}

\begin{figure*}
    \centering
    \includegraphics[width=\linewidth]{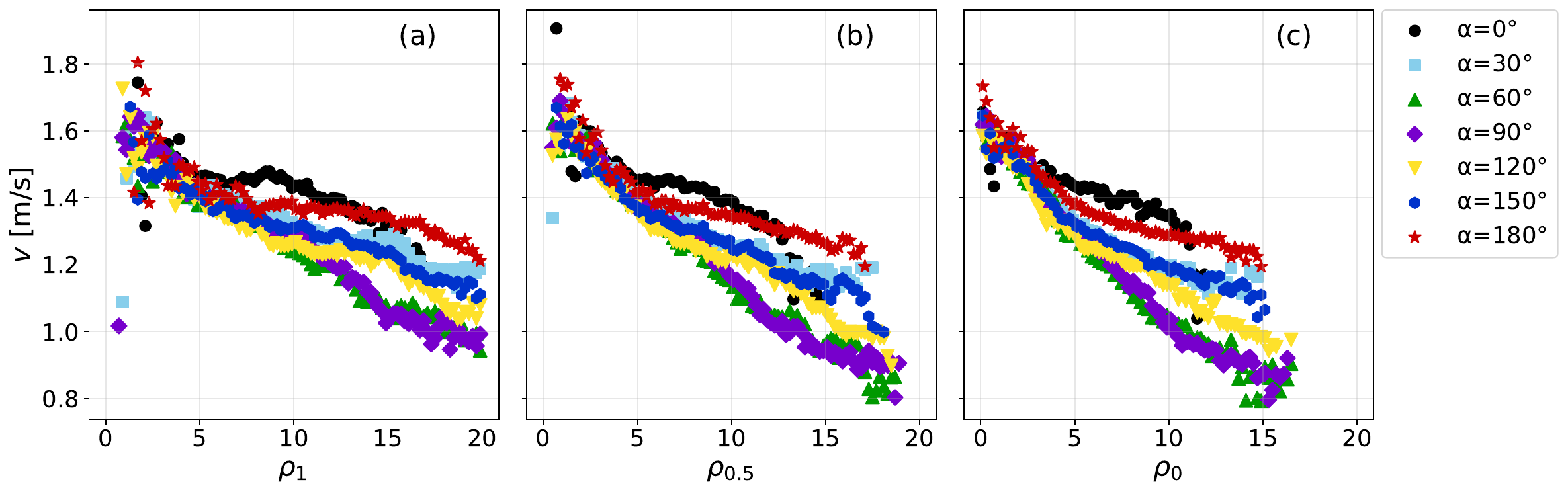}
    \caption{Velocity-crowdedness fundamental diagrams for each crossing angle $\alpha$ in our data set. The points in the plot are median values of observed velocity for a crowdedness bin of width 0.2 m$^{-2}$.}
    \label{fig:fig2}
\end{figure*}

Figure \ref{fig:fig1} demonstrates this robustness at the level of the temporal evolution of crowdedness. For all crossing angles except $\alpha=0\degree$, crowdedness increases during the interaction phase, reaches a maximum near the midpoint of the crossing process, and subsequently decreases as the two groups separate, consistent with previous observations based on Voronoi-cell densities \citep{mullick2025classifying}. Decreasing $c$ mainly reduces the magnitude of crowdedness, while the temporal evolution and the ordering across crossing angles remain largely unaffected.

Figure \ref{fig:fig2} shows that this robustness extends to the velocity-crowdedness fundamental diagrams as well. In all cases, velocity decreases with increasing crowdedness, and the relative ordering of the crossing angles is preserved. Lowering $c$ leads primarily to a horizontal shift of the curves toward smaller crowdedness values, while their overall shapes remain essentially unchanged. Thus, the geometry of the crossing interaction continues to determine the observed velocity-crowdedness relationship irrespective of the particular perception model employed.

\begin{figure*}
    \centering
    \includegraphics[width=\linewidth]{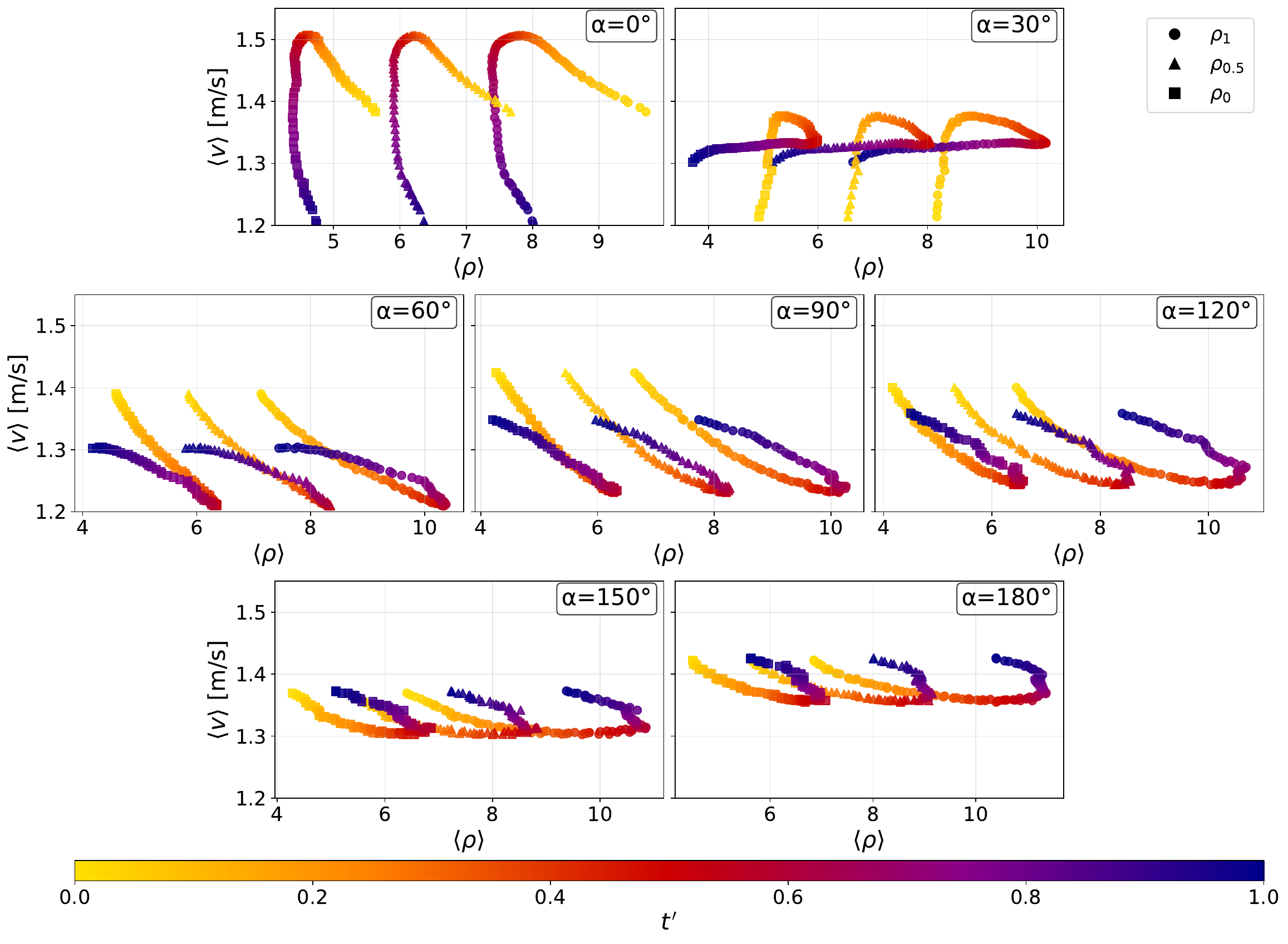}
    \caption{Dynamic velocity--crowdedness fundamental diagrams for different crossing angles $\alpha$. Each point corresponds to the average velocity $\langle v\rangle$ and crowdedness $\langle \rho_c \rangle$ of pedestrians at a given scaled time $t^\prime$ during a crossing-flow interaction. The color gradient represents the temporal evolution of the velocity-crowdedness relationship averaged across all trials.}
    \label{fig:fig3}
\end{figure*}

Figure \ref{fig:fig3} provides further evidence for this conclusion through the temporal evolution of the velocity-crowdedness phase space, which may also be interpreted as dynamic fundamental diagrams. The trajectories corresponding to $\rho_1$, $\rho_{0.5}$ and $\rho_0$ follow nearly identical paths, differing mainly by a horizontal displacement along the crowdedness axis. The overall shape of the trajectories and their temporal progression are preserved across all three perception models, indicating that perceptual anisotropy modifies the numerical value of crowdedness without significantly affecting the underlying pedestrian motion.

The robustness observed across Figures \ref{fig:fig1}-\ref{fig:fig3} is not obvious \textit{a priori}. Since pedestrians interact anisotropically and primarily respond to individuals within their field of view, one might expect that progressively discounting agents outside the field of view would substantially alter the local crowdedness experienced by pedestrians and, consequently, modify the associated dynamical relationships. Instead, we consistently observe that the qualitative behavior remains preserved across different perception models. These findings suggest that the collective organization of crossing flows is governed predominantly by the interaction geometry of the system, while perceptual anisotropy acts mainly as a rescaling of the crowdedness measure.

\section{Directional deviations: pedestrian adaptation}\label{sec:deviation}

To further characterize the microscopic dynamics of pedestrians during crossing interactions, we analyze the deviations of individual walking directions from their expected trajectories. Directional deviations provide information about how pedestrians adapt their motion in response to local interactions and congestion. Quantifying these deviations can therefore help us to understand the mechanisms underlying collision avoidance and the collective organization of crossing flows.

\begin{figure*}
    \centering
    \includegraphics[width=0.8\linewidth]{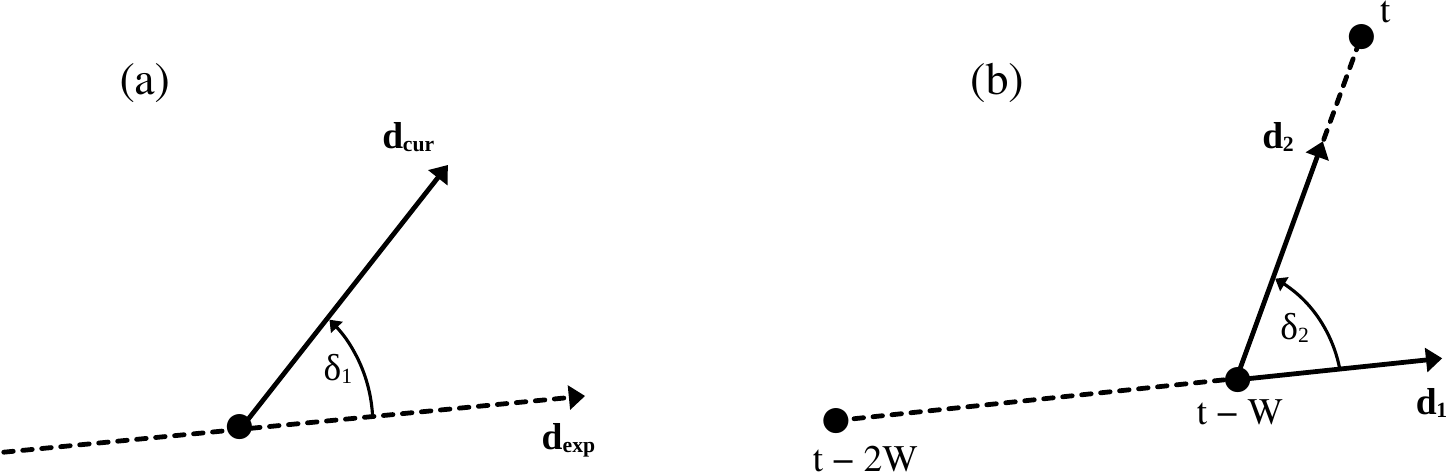}
    \caption{Schematic representation of the directional deviation measures used in this work. (a) The angle $\delta_1$ denotes the deviation of a pedestrian's current direction of motion, $\mathbf{d}_{\mathrm{cur}}$, from the expected direction of motion of its group, $\mathbf{d}_{\mathrm{exp}}$. (b) The angle $\delta_2$ represents the turning angle between two consecutive displacement vectors, $\mathbf{d}_1$ and $\mathbf{d}_2$, evaluated over time windows of duration $W=0.5$ s.
}
    \label{fig:fig4}
\end{figure*}

Two complementary measures of directional deviation are considered, denoted by $\delta_1$ and $\delta_2$, and illustrated schematically in Figure \ref{fig:fig4}. The quantity $\delta_1$ characterizes the instantaneous deviation of a pedestrian from the mean direction of motion of the group to which the pedestrian belongs. Specifically, if $\mathbf{d}_{\mathrm{cur}}$ denotes the current direction of motion of the pedestrian and $\mathbf{d}_{\mathrm{exp}}$ denotes the expected direction of motion of the corresponding group, then $\delta_1$ is defined as the angle between these two vectors, as shown in Figure \ref{fig:fig4}(a). Thus, $\delta_1$ measures the extent to which an individual deviates from the collective motion of its group at a given instant.

The second quantity, $\delta_2$, measures the temporal change in an individual's direction of motion. Let $\mathbf{d}_1$ and $\mathbf{d}_2$ denote the displacement vectors of a pedestrian computed over two consecutive time windows of duration $W=0.5$ s. The angle between these two vectors is defined as $\delta_2$, which is shown in Figure \ref{fig:fig4}(b). Consequently, $\delta_2$ quantifies the turning behavior of pedestrians, i.e., the extent to which an individual changes its walking direction between successive displacement windows.

Both $\delta_1$ and $\delta_2$ are computed for every pedestrian at every time step of the trajectory data. The two deviation measures are only weakly correlated across all crossing angles (corr. coeff. $r\lesssim 0.3$), despite the correlations being statistically significant ($p<10^{-5}$). This weak association indicates that $\delta_1$ and $\delta_2$ capture largely distinct aspects of pedestrian motion and can therefore be analyzed separately.

\begin{figure*}
    \centering
    \includegraphics[width=\linewidth]{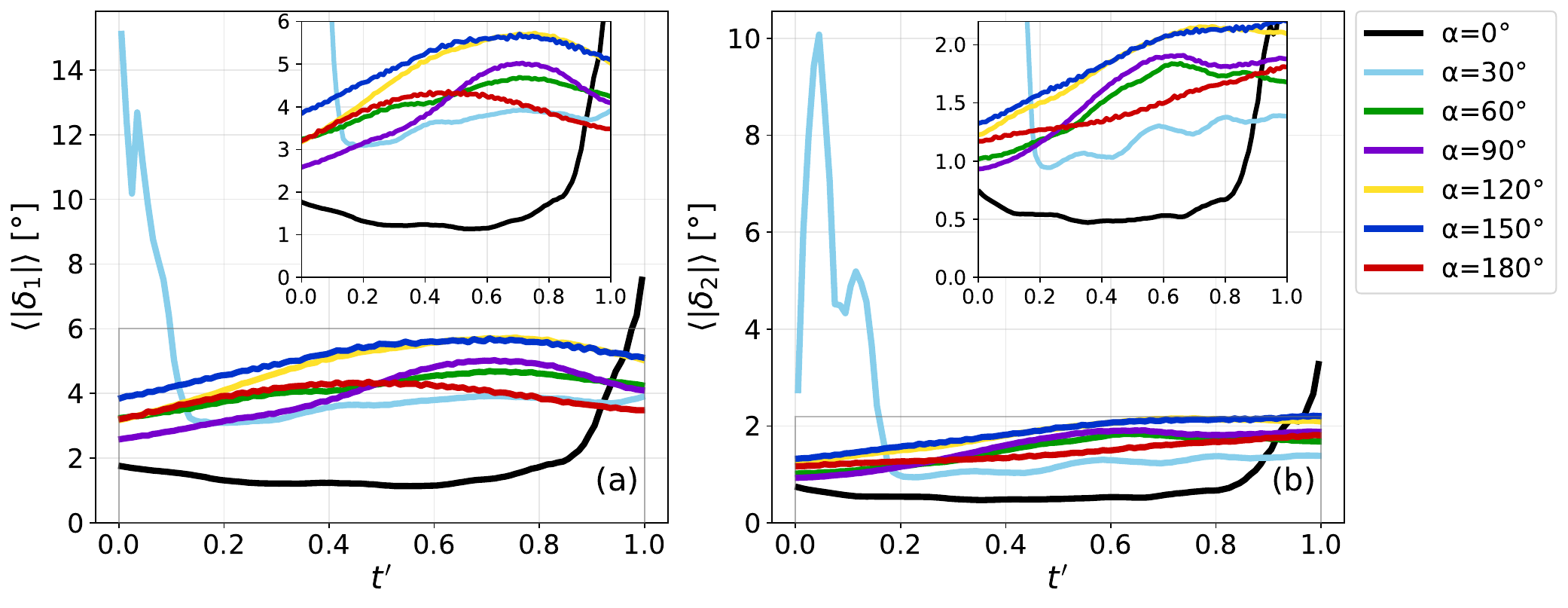}
    \caption{Temporal evolution of the mean absolute directional deviations (a) $\langle |\delta_1| \rangle$ and (b) $\langle |\delta_2| \rangle$ for different crossing angles $\alpha$. The quantity $\delta_1$ measures the instantaneous deviation from the expected direction of motion of the pedestrian's group, while $\delta_2$ quantifies the change in walking direction between successive displacement windows of duration $W=0.5$ s. The averages are computed over all pedestrians and all trials at each scaled time $t^\prime$.}
    \label{fig:fig5}
\end{figure*}

In Figure \ref{fig:fig5} we show the temporal evolution of the mean absolute directional deviations $\langle |\delta_1| \rangle$ and $\langle |\delta_2| \rangle$. Let us first discuss the results for $\delta_1$. For $\alpha=0\degree$, $\delta_1$ is the smallest during most of the interaction. This is expected because there is no conflict between the two groups, so pedestrians largely continue along their prescribed direction. For other crossing angles, $\delta_1$ increases during the interaction, reaches a broad maximum around $t^\prime=0.6$ to $0.8$ and then decreases. This is quite intuitive as the pedestrians deviate most strongly from their preferred directions when the overlap between the two streams is largest. The differences across crossing angles are surprisingly small, about $3\degree$ to 6$\degree$. Even for $\alpha=90\degree$ or $150\degree$, pedestrians do not need to deviate very much from their intended directions. This suggests that collision avoidance is achieved through many small adjustments rather than large detours.

The results for $\delta_2$ are undoubtedly more interesting. For all crossing angles except $0\degree$, $\delta_2$ remains remarkably small ($\lesssim 2\degree$) throughout interaction zone. This means that pedestrians do not perform abrupt turns while traversing the crossing region. Instead, they continuously adapt their trajectories through gradual corrections. Therefore, pedestrians do deviate from their intended directions ($\delta_1$ is finite), but they do so smoothly ($\delta_2$ remains small).

A notable exception is observed for $\alpha=30^\circ$, for which both deviation measures exhibit pronounced peaks at early times before rapidly approaching values comparable to those of the other crossing angles. This transient behavior appears to correspond to the transitional nature of the shallow crossing geometry between nearly-parallel flow and more strongly intersecting crossing flows, as further discussed in Appendix \ref{appendix:dev30}.

\begin{figure*}
    \centering
    \includegraphics[width=\linewidth]{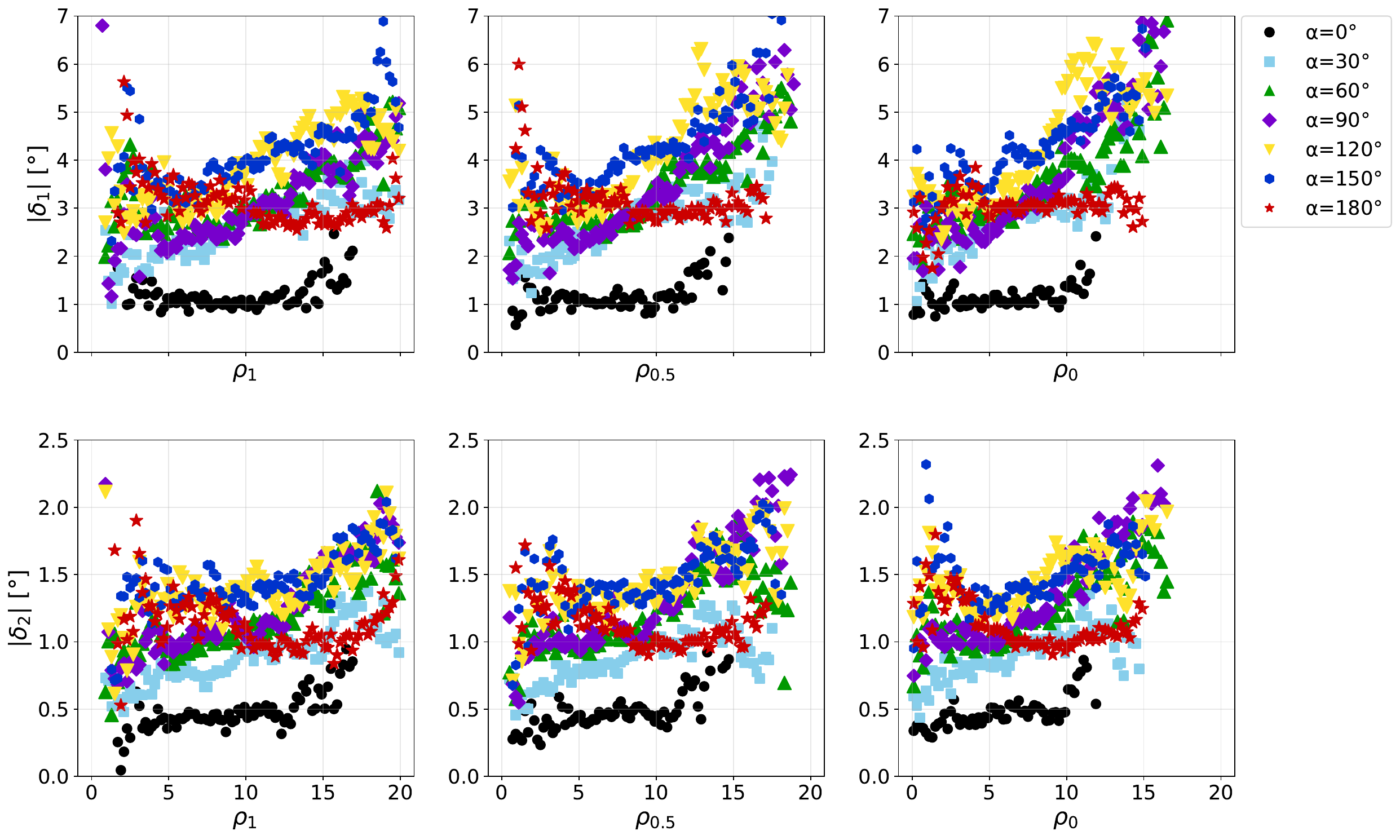}
    \caption{Deviations $|\delta_1|$ (top row) and $|\delta_2|$ (bottom row) as functions of crowdedness for different crossing angles $\alpha$. The three columns correspond to the indices of crowdedness $\rho_1$, $\rho_{0.5}$, and $\rho_0$, respectively, depending on the levels of perceptual anisotropy. The points in the plot are median values of observed deviation for a crowdedness bin of width 0.2 m$^{-2}$.}
    \label{fig:fig6}
\end{figure*}

Figure \ref{fig:fig6} shows the dependence of the absolute directional deviations $|\delta_1|$ and $|\delta_2|$ on crowdedness for different crossing angles and perception models. The case $\alpha=0^\circ$ serves as a reference corresponding to a unidirectional flow without crossing conflicts. In this case, both deviations remain small over a broad range of crowdedness, indicating that pedestrians are able to maintain their intended walking directions with minimal adjustments. Only at the largest crowdedness values do the deviations increase noticeably, reflecting the growing influence of local interactions in dense conditions.

For the crossing flows ($\alpha>0^\circ$), both $|\delta_1|$ and $|\delta_2|$ generally increase with increasing crowdedness, indicating that pedestrians deviate more strongly from their preferred directions and perform larger directional adjustments in denser regions. The increase, however, is nonlinear. For several crossing angles, particularly $\alpha=60^\circ$, $90^\circ$, $120^\circ$, and $150^\circ$, the deviations remain nearly constant at low crowdedness and increase rapidly only beyond a threshold value of crowdedness. This behavior suggests that pedestrians are able to preserve their intended trajectories in sparse conditions, whereas denser environments require progressively stronger avoidance maneuvers.

The magnitude of the two deviations differs substantially. While $|\delta_1|$, which measures the deviation from the expected group direction, can attain values of $5^\circ$--$6^\circ$, the turning angle $|\delta_2|$ remains comparatively small, typically below $2^\circ$ even at high crowdedness. Thus, pedestrians deviate appreciably from their preferred directions, but do not do so through abrupt turns. Instead, the trajectories are adjusted gradually through a sequence of small directional corrections. This observation points toward a high degree of collective organization in the crowd: rather than reacting independently through sharp avoidance maneuvers, pedestrians appear to coordinate their motion in a way that minimizes sudden changes in direction.

The results also reveal notable differences across crossing angles. Interestingly, the largest deviations are observed for oblique crossings ($\alpha=120^\circ$ and $150^\circ$), whereas the counterflow case ($\alpha=180^\circ$) exhibits comparatively small and almost density-independent deviations. This suggests that pedestrians in counterflows are able to self-organize into coherent lanes or stripe-like structures that facilitate motion while requiring only minor adjustments of direction. In contrast, oblique crossing geometries appear to impose stronger geometric constraints, leading to larger deviations from the preferred walking direction. The qualitative behavior remains remarkably similar across the three crowdedness definitions $\rho_1$, $\rho_{0.5}$, and $\rho_0$, indicating once again that perceptual anisotropy primarily rescales the crowdedness measure without fundamentally altering the underlying deviation-crowdedness relationships.

The temporal evolution of the velocity-deviation phase space shows loop-like trajectories for most crossing angles, indicating that the disruption and velocity recovery phases do not follow the same dynamical pathway. The resulting dependence suggests that the directional state of the pedestrians cannot be inferred from their instantaneous velocity alone (see Appendix \ref{appendix:velo-dev}). Similarly, the deviation-crowdedness phase spaces exhibit non-retracing, loop-like trajectories, indicating that the same crowdedness can correspond to different directional states during the build-up and relaxation phases of the interaction. The loops are more pronounced for $\langle|\delta_1|\rangle$ than for $\langle|\delta_2|\rangle$, suggesting a stronger memory effect in the accumulated deviation from the expected group direction than in incremental turning adjustments. Their qualitative structure remains robust across the three perception models, with perceptual anisotropy producing primarily a quantitative rescaling of crowdedness (see Appendix \ref{appendix:dev-crowd}).

\section{Acceleration and speed-adjustment dynamics}\label{sec:accln}

While the temporal evolution of velocity characterizes the overall slowing down and subsequent recovery of pedestrians during a crossing interaction, acceleration $a$ provides a more direct measure of how pedestrians adapt their motion as the interaction develops. In particular, the sign of the acceleration distinguishes between phases of deceleration and recovery, while its magnitude denotes the rate at which pedestrians modify their walking speed. Studying acceleration therefore complements the velocity analysis by revealing when, and how strongly, pedestrians respond to the evolving crossing configuration.

\begin{figure*}[h!]
    \centering
    \includegraphics[width=\linewidth]{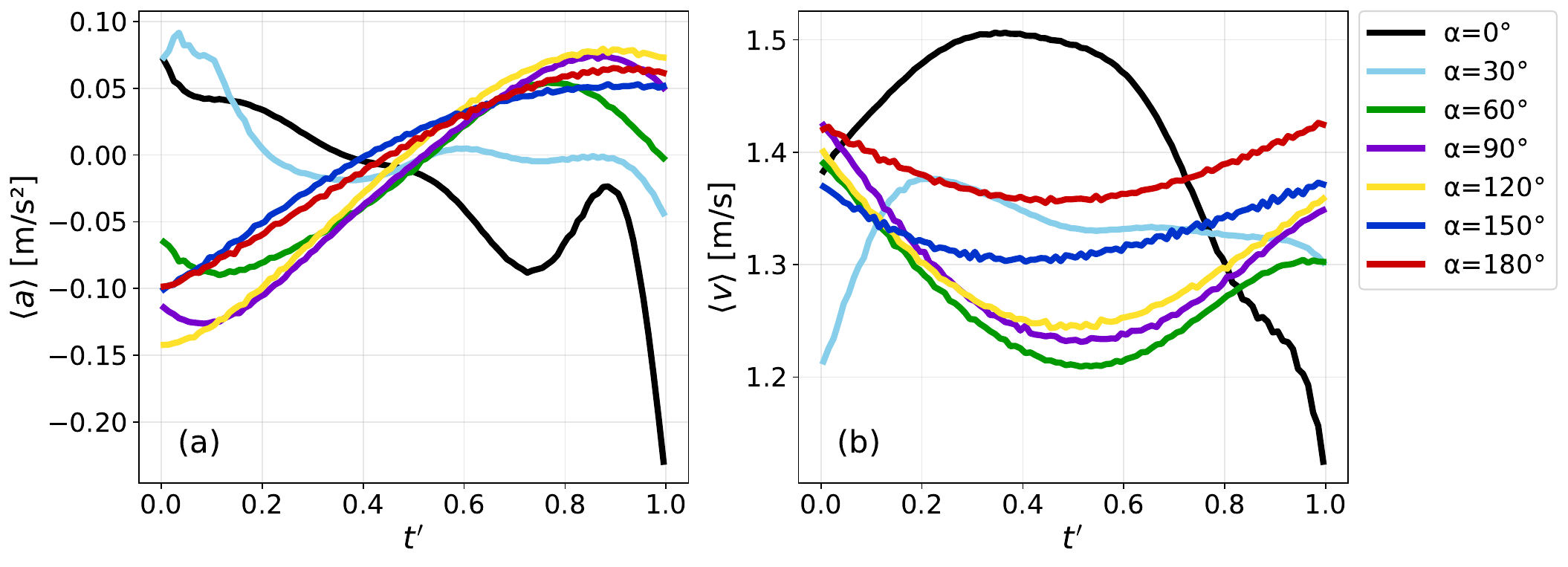}
    \caption{Temporal evolution of (a) mean acceleration $\langle a\rangle$ and (b) mean velocity $\langle v\rangle$ as functions of the scaled time $t^\prime$ for different crossing angles $\alpha$. The averages are computed over all pedestrians and all trials at each scaled time. The plot of time dependence for mean velocity was also shown in \cite{mullick2025classifying}.}
    \label{fig:fig7}
\end{figure*}

The temporal evolution of the mean acceleration $\langle a\rangle$ is shown in Figure \ref{fig:fig7}(a). For most crossing flows ($\alpha>0^\circ$), the pedestrians initially show negative acceleration, corresponding to the decrease in velocity observed during the early stage of the interaction. As the two groups increasingly overlap, the magnitude of the deceleration gradually decreases, and the acceleration approaches zero around the middle of the crossing process, approximately when the velocity reaches its minimum, as shown in Figure \ref{fig:fig7}(b). Thereafter, the acceleration becomes positive as the groups begin to separate and pedestrians recover their walking speeds. The acceleration curves therefore provide a dynamical counterpart to the characteristic U-shaped velocity curves: the initial deceleration produces the descending branch of the velocity curve, the zero crossing of the acceleration approximately identifies the velocity minimum, and the subsequent positive acceleration corresponds to the recovery phase.

The magnitude and timing of these changes also depend on the crossing angle. The $\alpha=90^\circ$ and $120^\circ$ cases exhibit some of the strongest initial deceleration, consistent with the pronounced reduction in velocity observed for these crossing geometries, followed by relatively strong positive acceleration during the recovery phase. The $\alpha=150^\circ$ and $180^\circ$ cases show weaker initial deceleration, in accordance with their comparatively lower reduction in velocity. The $\alpha=30^\circ$ case exhibits a distinct early-time behavior, with an initial positive acceleration followed by deceleration and only a weak recovery at later times. This shows the corresponding non-monotonic early evolution of its velocity and distinguishes the shallow-angle crossing from the other crossing configurations. The $\alpha=60^\circ$ case also differs slightly at late times, with the acceleration decreasing again after reaching positive values, consistent with the flattening of its velocity curve toward the end of the interaction.

On the other hand, the $\alpha=0^\circ$ case displays a fundamentally different temporal evolution, as expected in the absence of a crossing conflict. The initially positive acceleration corresponds to the increase in velocity during the first part of the observation period, after which the acceleration becomes negative as the velocity reaches a maximum and begins to decrease. However, the pronounced decrease in both velocity and acceleration toward the end of the normalized time interval should not be interpreted as an interaction-induced effect. Since no crossing conflict occurs for $\alpha=0^\circ$, this terminal behavior is most likely associated with entrance and exit effects arising from the finite observation region. The temporal variation observed for $\alpha=0^\circ$ should therefore be regarded as a baseline influenced by the experimental geometry rather than as a signature of crossing-flow dynamics.

Taken together, we see an interesting asymmetry between the deceleration and recovery phases of the crossing interaction. The magnitude of the initial deceleration varies considerably across crossing angles and is broadly consistent with the corresponding loss of velocity, indicating that crossing geometry strongly influences the extent to which pedestrians are forced to slow down. In contrast, once the interaction is largely resolved, the positive accelerations associated with speed recovery converge to a much narrower range across crossing angles. Thus, while the geometry of the crossing strongly determines the disruption experienced during the interaction, the subsequent recovery toward the desired walking speed appears to be comparatively similar across different crossing configurations.

\begin{figure*}
    \centering
    \includegraphics[width=\linewidth]{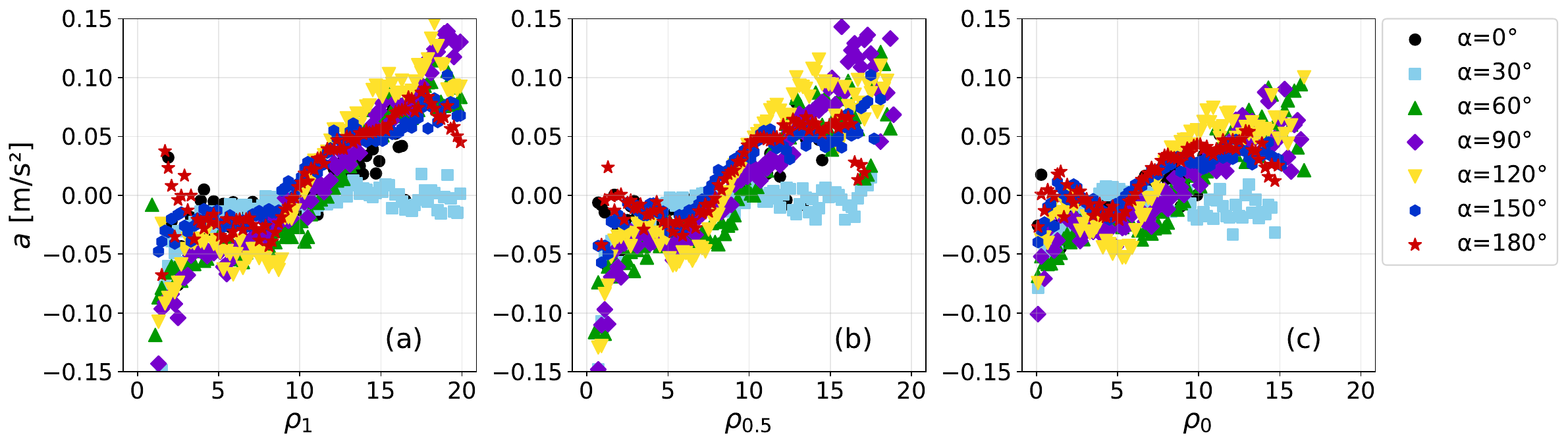}
    \caption{Acceleration-crowdedness relationships for different crossing angles $\alpha$ and indices of crowdedness (a) $\rho_1$, (b) $\rho_{0.5}$, and (c) $\rho_0$, depending on three levels of perceptual anisotropy. Each point corresponds to the median acceleration for a crowdedness bin of width 0.2 m$^{-2}$.}
    \label{fig:fig8}
\end{figure*}

Figure \ref{fig:fig8} shows the acceleration--crowdedness relationships for different crossing angles and perception models. Unlike a conventional fundamental diagram, this plot does not represent a unique state relation between acceleration and crowdedness; rather, crowdedness acts partly as a proxy for the stage of the crossing interaction, with the negative and positive-acceleration branches corresponding broadly to the deceleration and recovery phases, respectively.

For most crossing angles, acceleration is negative at low-to-intermediate crowdedness and becomes positive at higher crowdedness. This behavior can be understood from the temporal evolution of the crossing process. During the initial phase, the two groups increasingly overlap, crowdedness rises, and pedestrians decelerate. As the interaction progresses, the deceleration gradually weakens and the acceleration changes sign, after which pedestrians begin to recover their walking speeds. The acceleration--crowdedness relationship therefore distinguishes the deceleration and recovery phases of the interaction more directly than the corresponding velocity--crowdedness fundamental diagram.

The magnitude of the acceleration response depends on the crossing angle. The intermediate crossing angles, particularly $\alpha=60^\circ$, $90^\circ$, and $120^\circ$, exhibit relatively strong deceleration at lower crowdedness and some of the largest positive accelerations at higher crowdedness, consistent with their pronounced velocity loss and subsequent recovery. In comparison, the acceleration response is generally weaker for the larger crossing angles $\alpha=150^\circ$ and $180^\circ$. The $\alpha=30^\circ$ case is distinct, with acceleration remaining close to zero over much of the crowdedness range, consistent with its atypical temporal velocity evolution. %The $\alpha=0^\circ$ case should be interpreted separately, since no crossing conflict occurs and its temporal dynamics are influenced by entrance and exit effects.

The qualitative structure of these relationships remains largely unchanged across $\rho_1$, $\rho_{0.5}$, and $\rho_0$. Decreasing the anisotropy parameter $c$ primarily shifts the data toward lower crowdedness values, while the angle-dependent patterns and the transition from negative to positive acceleration remain broadly preserved. This is consistent with the velocity-crowdedness (Figure \ref{fig:fig2}) and deviation-crowdedness (Figure \ref{fig:fig6}) relationships discussed earlier, further indicating that perceptual anisotropy mainly rescales the crowdedness measure without fundamentally altering the underlying crossing dynamics.

\section{Implications for pedestrian flow dynamics}\label{sec:implications}

Our results obtained across crowdedness, velocity, directional deviation, and acceleration provide several broader implications for the characterization of pedestrian interactions. In particular, they highlight the importance of temporal evolution, complementary behavioral variables, the representation of local interaction state, and interaction geometry in describing the organization of pedestrian flows.

\textit{Dynamic characterization of pedestrian interactions}: The results show that pedestrian interactions cannot always be characterized uniquely by instantaneous measures of crowdedness or motion. The non-retracing trajectories observed in the velocity-crowdedness (Figure \ref{fig:fig1}), velocity–deviation (Appendix \ref{appendix:velo-dev}) and deviation–crowdedness phase spaces (Appendix \ref{appendix:dev-crowd}) indicate that similar instantaneous values can correspond to different behavioral states depending on the preceding evolution of the interaction. Similarly, the acceleration–crowdedness relationship (Figure \ref{fig:fig8}) separates the initial deceleration from the subsequent speed-recovery phase, even when comparable crowdedness levels are observed. More broadly, the results point to path dependence as a recurring feature of transient pedestrian traffic, suggesting that conventional single-valued fundamental relationships may not fully characterize the flow. Whereas dynamic fundamental diagrams \cite{mullick2025classifying} that incorporate temporal evolution can reveal information about the recent interaction history that is unavailable from isolated instantaneous observations. Conventional fundamental diagrams show the states; dynamic fundamental diagrams show the process.

%\textit{Dynamic rather than static crowd assessment}: The results suggest that crowd conditions should be interpreted dynamically rather than solely through instantaneous measures of density or crowdedness. The non-retracing trajectories observed in the velocity–deviation (Figure 2 in Appendix \ref{supmat}) and deviation–crowdedness (Figures 3 and 4 in Appendix \ref{supmat}) phase spaces show that the same instantaneous crowdedness or velocity can correspond to different behavioral states depending on whether the interaction begins or ends. Similarly, the acceleration–crowdedness (Figure \ref{fig:fig8}) relationship distinguishes the initial deceleration phase from the subsequent recovery phase, even when comparable crowdedness levels are observed. Thus, static thresholds based on a single variable may not fully characterize the state of an interacting crowd; incorporating the recent evolution of motion variables may help distinguish a crowd undergoing increasing disruption from one already recovering from an interaction.

\textit{Complementary measures of behavioral adaptation}: The directional-deviation results (section \ref{sec:deviation}) further demonstrate the importance of characterizing pedestrian adaptation through multiple behavioral variables. Although pedestrians deviate appreciably from the expected directions of their groups, the changes between successive walking directions generally remain small, showing that these deviations emerge predominantly through sequences of gradual corrections rather than abrupt turns. The weak correlation between $\delta_1$ and $\delta_2$ (all $r<0.3$, $p<10^{-5}$) further indicates that departure from the expected group direction and incremental turning represents distinct aspects of pedestrian motion. These observations suggest that not only the magnitude of directional deviation, but also the manner in which such deviation occurs, may provide information about the organization of pedestrian interactions. In particular, a transition from smooth, incremental adjustments toward more frequent or abrupt directional changes could potentially signal a deterioration of collective organization, although this concept requires validation in more congested or safety-critical conditions.

%\textit{Behavioral indicators of organized adjustment}: The directional-deviation results (section \ref{sec:deviation}) further suggest that behavioral response variables can complement conventional measures based on crowdedness and speed. Although pedestrians deviate from their expected group directions during crossing interactions, the turning angle $\delta_2$ between successive displacement windows generally remains small, indicating that collision avoidance is achieved predominantly through sequences of gradual directional corrections rather than abrupt maneuvers. The weak correlation (all $r<0.3$, $p<10^{-5}$) between the two deviation measures also shows that deviation from the expected direction and incremental turning capture distinct aspects of pedestrian behavior. These observations suggest that not only the magnitude of directional deviation, but also the manner in which such deviation occurs, may provide information about the organization of pedestrian interactions. In particular, a transition from smooth, incremental adjustments toward more frequent or abrupt directional changes could potentially signal a deterioration of collective organization, although this possibility requires validation in more congested or safety-critical conditions.

\textit{Robustness to the representation of perceptual anisotropy}: A further implication concerns how the local interaction state is represented in empirical analyses of pedestrian motion. Across the temporal profiles, velocity–crowdedness diagrams (Figure \ref{fig:fig1}), deviation–crowdedness (Figure \ref{fig:fig6}) and acceleration–crowdedness relationships (Figure \ref{fig:fig8}), and the corresponding phase-space trajectories (Figure \ref{fig:fig3} and Appendices \ref{appendix:velo-dev} \& \ref{appendix:dev-crowd}), varying the anisotropy parameter $c$ changes primarily the numerical scale of crowdedness while leaving the qualitative dynamics and crossing-angle dependence largely unchanged, quite similar to \cite{WANG2023104400}. Thus, for the crossing flows studied here, the inferred geometry-dependent behavioral relationships are robust to the precise weighting assigned to pedestrians outside the field of view. This robustness suggests that the principal dynamical conclusions are not an artefact of a particular parametrization of perceptual anisotropy. Although whether the same holds for bottlenecks, restricted-visibility environments, or strongly heterogeneous flows needs to be thoroughly investigated, as was done in \cite{DUIVES2015162}.

%\textit{Robustness to the perception model}: A further practical consideration concerns the definition of local crowdedness. Across the temporal profiles, velocity–crowdedness diagrams (Figure \ref{fig:fig1}), deviation–crowdedness relationships (Figure \ref{fig:fig6}), acceleration–crowdedness relationships (Figure \ref{fig:fig8}), and the corresponding phase-space trajectories (Figure \ref{fig:fig3} and Figures 2, 3 \& 4 in Appendix \ref{supmat}), varying the anisotropy parameter $c$ changes primarily the numerical scale of crowdedness while leaving the qualitative dynamics and crossing-angle dependence largely unchanged. For the crossing flows studied here, this suggests that safety-oriented assessments based on local crowdedness may be relatively robust to the precise weighting assigned to pedestrians outside the field of view. At the same time, this result should not be generalized to other crowd configurations without further investigation, particularly situations involving bottlenecks, restricted visibility, or strongly heterogeneous motion, where perceptual anisotropy may play a more substantial role.

\textit{Interaction geometry as an organizing variable}: Finally, the results emphasize the central role of interaction geometry in shaping pedestrian flow dynamics. Crossing angle systematically influences velocity loss, deceleration, and directional deviation, but the observed responses do not follow a simple monotonic ordering with angle, as also shown in \cite{mullick2025classifying}. Intermediate and oblique crossings produce pronounced disruption and directional adaptation, counterflow exhibits comparatively small directional deviations consistent with efficient self-organization, and the $30\degree$ case displays distinct transient dynamics associated with its position between nearly-parallel and more strongly intersecting flows. Interaction geometry therefore appears to organize not only the magnitude of pedestrian response, but also the pathway through which the flow adapts over time. Characterizing interacting pedestrian streams consequently requires explicit consideration of their relative directions of motion rather than treating crowdedness or other local state variables independently of the underlying flow configuration.

%\textit{Geometry-specific interpretation of crowd behavior}: Our results emphasize that behavioral indicators should be interpreted in relation to the geometry of the interacting flows. Crossing angle systematically influences velocity loss, deceleration, and directional deviation: intermediate and oblique crossings produce pronounced disruption, whereas counterflow can exhibit comparatively modest directional deviations, consistent with efficient self-organization. The $30\degree$ case further demonstrates that shallow crossing geometries may involve distinct transient dynamics that are not captured by a simple monotonic ordering with crossing angle. Consequently, a given level of crowdedness, speed reduction, or directional deviation need not have the same behavioral significance across different flow configurations. Crowd assessment in locations with intersecting pedestrian streams may therefore benefit from accounting explicitly for interaction geometry rather than applying identical interpretations or thresholds irrespective of the directions of movement.

\section{Conclusions}\label{sec:conclusions}

In this paper, we investigated how local crowdedness and behavioral adaptation evolve in pedestrian crossing flows and how the resulting dynamics depend on interaction geometry and perceptual anisotropy. We introduced a distance-weighted measure of local crowdedness and considered different levels of perceptual anisotropy by progressively reducing the contribution of pedestrians outside the focal pedestrian's field of view. Across the temporal evolution of crowdedness and its relationships with velocity, directional deviation, and acceleration, the principal effect of increasing perceptual anisotropy was a quantitative reduction in the estimated crowdedness. The qualitative dynamical patterns, their temporal progression, and their dependence on crossing angle remained largely preserved. This repeated robustness suggests that, for the crossing flows considered here, the collective dynamics are governed more strongly by the geometry of the interacting streams than by the precise perceptual weighting used to quantify local crowdedness.

The analysis of behavioral responses provides a complementary picture of how pedestrians adapt to these interactions. Pedestrians can deviate appreciably from the expected direction of their group, yet the changes between successive walking directions remain comparatively small, indicating that trajectory adaptation occurs predominantly through sequences of smooth, incremental corrections rather than abrupt turns. Such gradual directional adjustments may also provide the microscopic mechanism through which self-organized structures, such as stripes, emerge during crossing interactions, as coordinated patterns can develop through the accumulation of small individual trajectory corrections.

The acceleration dynamics further reveal an asymmetry between disruption and recovery: the magnitude of the initial deceleration and the associated velocity loss vary considerably with crossing angle, whereas the subsequent recovery accelerations converge to a comparatively narrower range. Moreover, the non-retracing trajectories observed in the different phase spaces show that instantaneous crowdedness or velocity does not uniquely determine the behavioral state of the flow. Similar instantaneous conditions may occur while an interaction is developing or resolving, highlighting the importance of the recent dynamical history of the crowd.

Taken together, these findings show that pedestrian crossing flows are better characterized through the dynamic fundamental diagrams compared to conventional fundamental diagrams based on instantaneous measurements. The crowdedness measure introduced here should be interpreted as an \textit{index of local interaction intensity} rather than as a conventional occupancy-based density, despite its inverse-area dimensionality. However, the field-of-view weighting represents directional sensitivity rather than a literal model of visual perception. Future work should examine whether the robustness observed here extends to qualitatively different flow configurations, where directional sensitivity may play a more crucial role.\\

\noindent\textbf{\large Funding}

This research was funded by the National Science Centre (NCN), Poland through SONATA (Grant 2022/47/D/HS4/02576).\\

\noindent\textbf{\large Declaration of competing interest}

The authors declare that they have no known competing financial interests or personal relationships that could have appeared to influence the work reported in this paper.\\

\noindent\textbf{\large Data availability}

The data set for experimental trials of crossing flows is publicly available at \url{https://doi.org/10.5281/zenodo.5718430}. The computational codes used for analyses presented in this paper could be found at \url{https://github.com/holowaczi/perception-based-crowdedness}.

%\bibliographystyle{cas-model2-names}

\bibliography{cas-refs}

\appendix
\section{Appendix: Interpretation of the initial deviation peaks for $\alpha=30\degree$}\label{appendix:dev30}

\begin{figure}[h!]
    \centering
    \includegraphics[width=\linewidth]{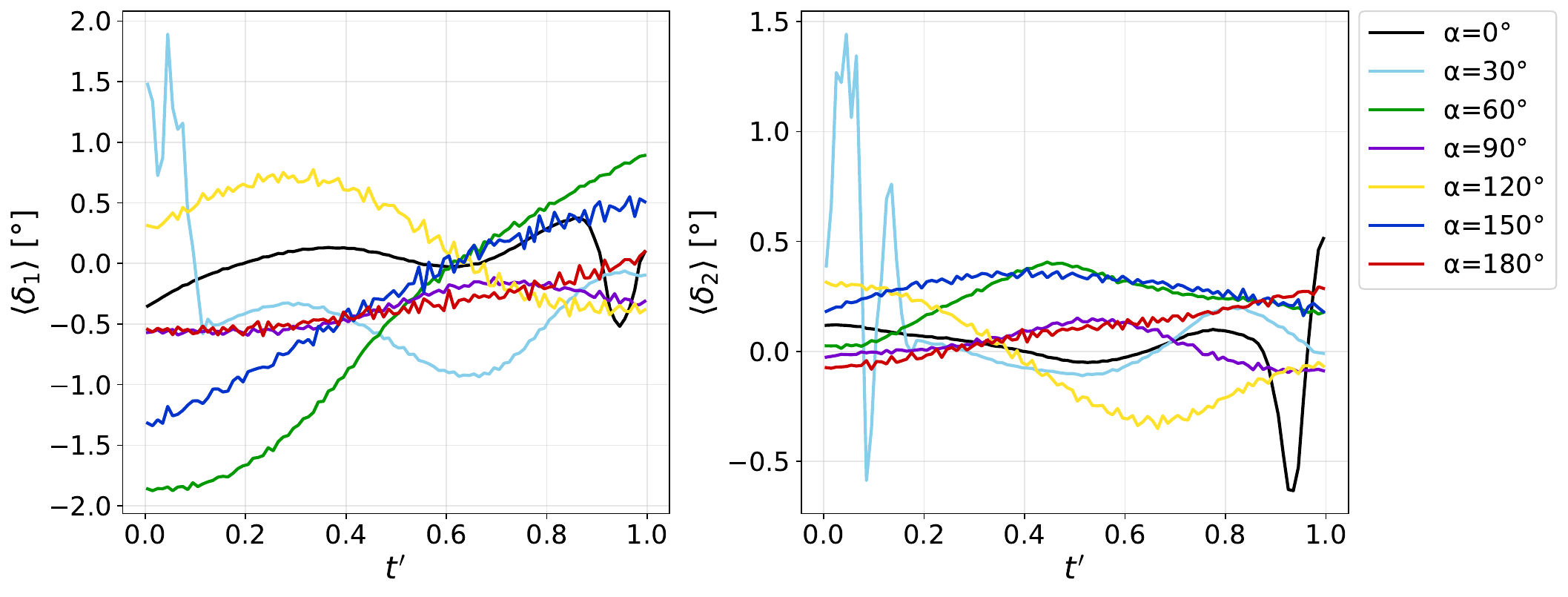}
    \caption{Temporal evolution of the mean signed directional deviations $\langle \delta_1 \rangle$ (left) and $\langle \delta_2 \rangle$ (right) for different crossing angles $\alpha$. The averages are computed over all pedestrians and all trials at each scaled time $t^\prime$.}
    \label{fig1}
\end{figure}

\noindent A distinct early-time behavior is observed for the shallow crossing angle $\alpha=30^\circ$. In the plots of the mean absolute deviations (shown in Figure 5 of the main text), both $\langle |\delta_1| \rangle$ and $\langle |\delta_2| \rangle$ exhibit pronounced peaks during the initial part of the interaction, approximately for $t^\prime \lesssim 0.15$. These peaks are much larger than those observed for the other crossing angles and therefore require a separate interpretation. Importantly, the corresponding signed averages (see Figure \ref{fig1}), $\langle \delta_1 \rangle$ and $\langle \delta_2 \rangle$, also show strong early-time excursions. This indicates that the peaks in the absolute deviations are not simply produced by random angular fluctuations of both signs, but rather reflect a coherent, preferentially directed reorientation of pedestrians during the early stage of the $30^\circ$ interaction.\\

\noindent This behavior can be understood by considering the geometry of the shallow crossing. At $\alpha=30^\circ$, the two groups possess a large common component of motion along nearly the same direction, while their transverse relative motion is comparatively weak. As a result, the initial configuration is closer to a nearly-parallel flow than to a strongly intersecting crossing flow. In this regime, pedestrians may initially move in a manner that retains some features of the $\alpha=0^\circ$ case, before the weak transverse component of the interaction becomes sufficiently important to generate a genuine crossing-flow configuration. The early peaks in $\langle |\delta_1| \rangle$ and $\langle |\delta_2| \rangle$ may therefore correspond to a transient reorganization phase, during which pedestrians adjust their walking directions in order to accommodate the emerging interaction between two nearly co-moving streams.\\

\noindent This interpretation is supported by the velocity and acceleration profiles (Figure 7). Unlike the larger crossing angles, which show an early decrease in velocity associated with the onset of the crossing conflict, the $\alpha=30^\circ$ case exhibits an initial increase in velocity and a corresponding positive acceleration. Only after this early phase does the acceleration become negative and the velocity begin to follow a pattern more consistent with the other crossing flows. Thus, the $30^\circ$ case appears to undergo a delayed transition from an initially nearly-parallel-flow-like state toward a more conventional crossing-flow state. The early directional deviations are therefore consistent with the dynamical adjustment required during this transition.\\

\noindent The later-time behavior further supports this interpretation. After the initial transient, the values of $\langle |\delta_1| \rangle$ and $\langle |\delta_2| \rangle$ for $\alpha=30^\circ$ decrease rapidly and become comparable to those of the other crossing angles. This suggests that the large early deviations are not a persistent property of shallow-angle crossings, but are associated with the initial establishment of the collective flow configuration. Once this reorganization has occurred, pedestrians in the $30^\circ$ case appear to adjust their trajectories through small, gradual corrections, similar to the behavior observed for the other crossing angles.\\

\noindent Overall, the $\alpha=30^\circ$ case may be interpreted as a transitional regime between unidirectional flow and more strongly intersecting crossing flows. Its early-time behavior retains some resemblance to the $\alpha=0^\circ$ case because of the large common component of motion, while its later-time dynamics approach those of the larger crossing angles as transverse interactions become more relevant. The unusual early peaks in directional deviation therefore provide additional evidence that shallow crossing angles involve a distinct reorganization process before a stable collective crossing pattern is established.\\

\section{Appendix: Temporal evolution of the velocity-deviation phase space}\label{appendix:velo-dev}

\noindent To further study how changes in walking speed are coupled to directional adjustments during the crossing process, we consider the joint temporal evolution of velocity and the two deviation measures. We plot these quantities in phase space to determine whether disruption and subsequent recovery follow the same dynamical pathway.\\

\begin{figure}[h!]
    \centering
    \includegraphics[width=\linewidth]{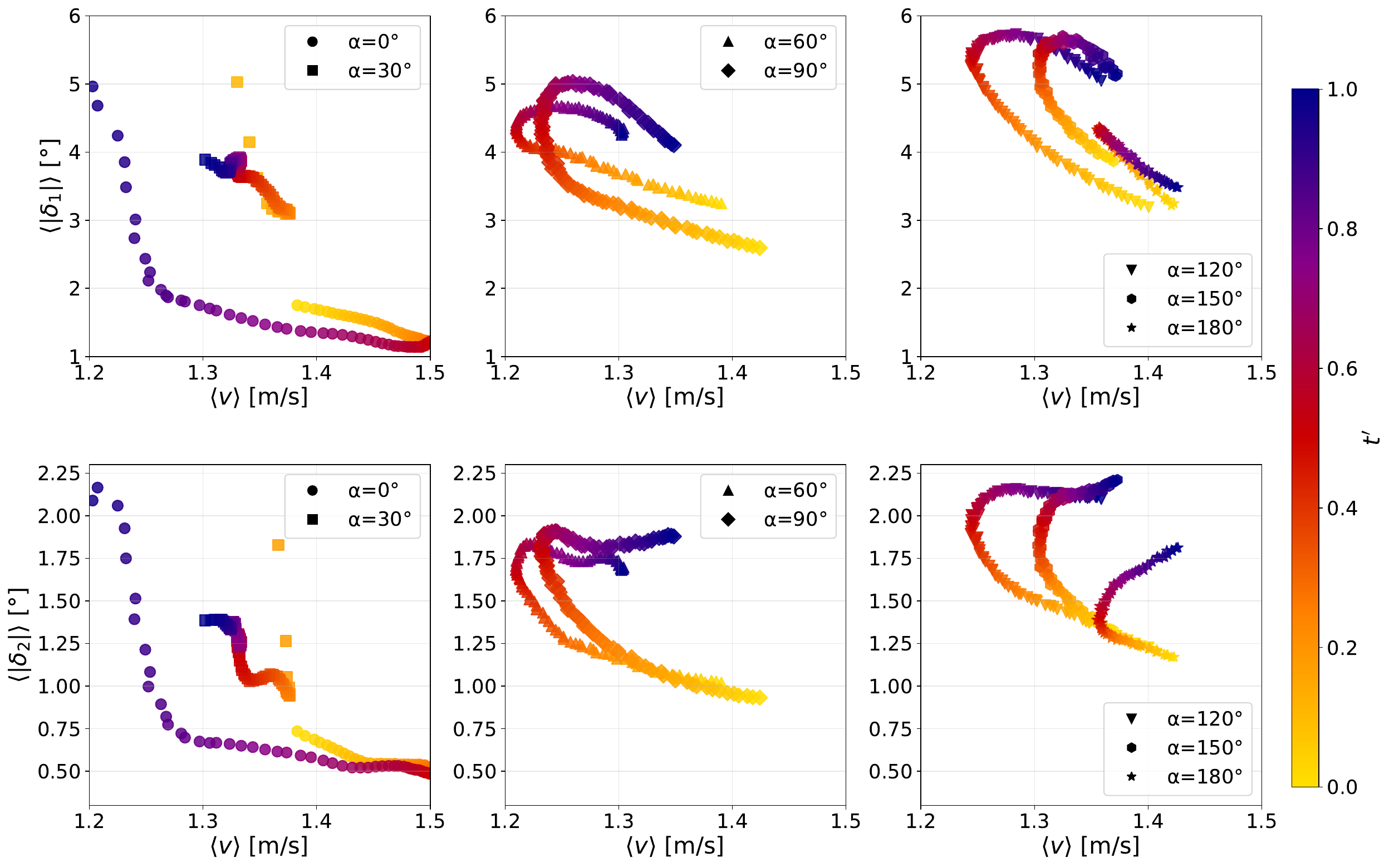}
    \caption{Dynamic velocity-deviation phase space for different crossing angles $\alpha$. Each point corresponds to the average velocity $\langle v\rangle$ and deviation $\langle \delta \rangle$ of pedestrians at a given scaled time $t^\prime$ during a crossing-flow interaction. The color gradient represents the temporal evolution of the velocity-deviation relationship averaged across all trials.}
    \label{fig2}
\end{figure}

\noindent Figure \ref{fig2} presents the temporal evolution of the velocity-deviation phase space for the two deviation measures, $\langle |\delta_1| \rangle$ and $\langle |\delta_2| \rangle$. For most crossing angles, the trajectories form curved or loop-like paths rather than collapsing onto a single relationship between velocity and directional deviation. Thus, the same mean velocity may be associated with different levels of deviation depending on the stage of the interaction. This indicates a history dependence of the velocity-deviation relationship: the directional state of the pedestrians cannot be inferred from their instantaneous velocity alone.\\

\noindent The loop-like structure can be related to the successive stages of the crossing process. During the initial interaction phase, pedestrians generally lose speed while their directional deviations increase as they adjust their trajectories to the opposing flow. At later times, the velocity recovers, but the deviations do not retrace the same path in reverse. Instead, the two quantities evolve at different rates, giving rise to the observed loops. This behavior is particularly evident for the intermediate and large crossing angles, whereas the $\alpha=0^\circ$ case follows a comparatively simple trajectory, consistent with the absence of a crossing conflict. The $\alpha=30^\circ$ case again exhibits a distinct evolution, reflecting the unusual early-time dynamics of this shallow crossing geometry discussed in the previous section.\\

\noindent The phase-space trajectories also reveal a systematic dependence on crossing angle. The smaller and intermediate crossing angles generally occupy lower deviation ranges, whereas $\alpha=120^\circ$, $150^\circ$, and $180^\circ$ reach substantially larger values of $\langle |\delta_1| \rangle$. Nevertheless, $\langle |\delta_2| \rangle$ remains considerably smaller than $\langle |\delta_1| \rangle$ across all crossing geometries. This reinforces the interpretation that pedestrians may progressively depart from their expected group directions during the interaction while changing their instantaneous walking directions through comparatively small, gradual adjustments. The non-retracing trajectories further suggest that the disruption and recovery phases involve different microscopic processes rather than constituting simple time reversals of one another.\\

\section{Appendix: Temporal evolution of the deviation-crowdedness phase space}\label{appendix:dev-crowd}

\begin{figure}[h!]
    \centering
    \includegraphics[width=\linewidth]{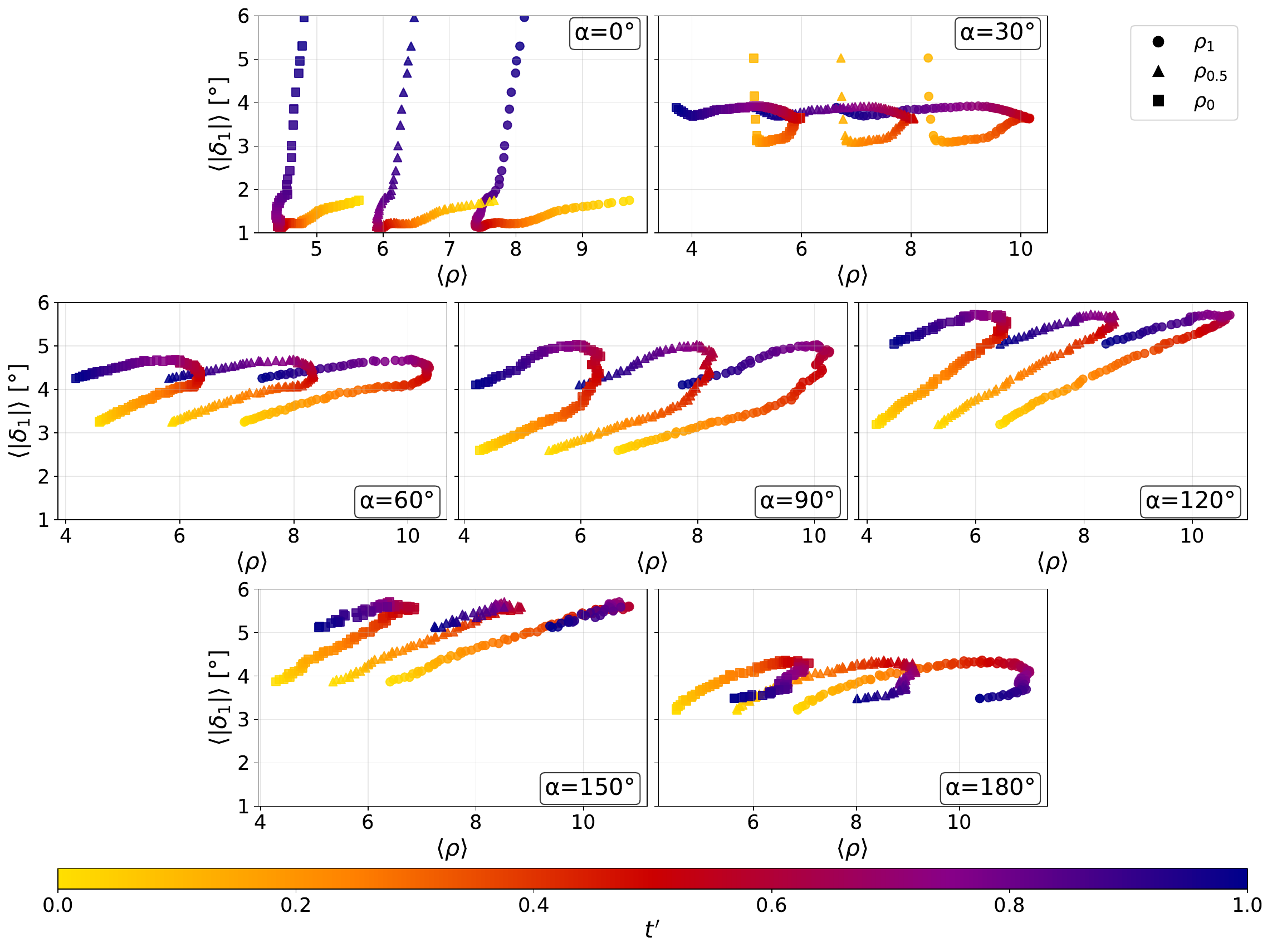}
    \caption{Temporal evolution of mean $|\delta_1|$ and crowdedness for different crossing angles $\alpha$. Each point corresponds to the average deviation $\langle |\delta_1|\rangle$ and crowdedness $\langle \rho \rangle$ of pedestrians at a given scaled time $t^\prime$ during a crossing-flow interaction. The color gradient represents the temporal evolution of the deviation-crowdedness relationship averaged across all trials.}
    \label{fig3}
\end{figure}

\begin{figure}[h!]
    \centering
    \includegraphics[width=\linewidth]{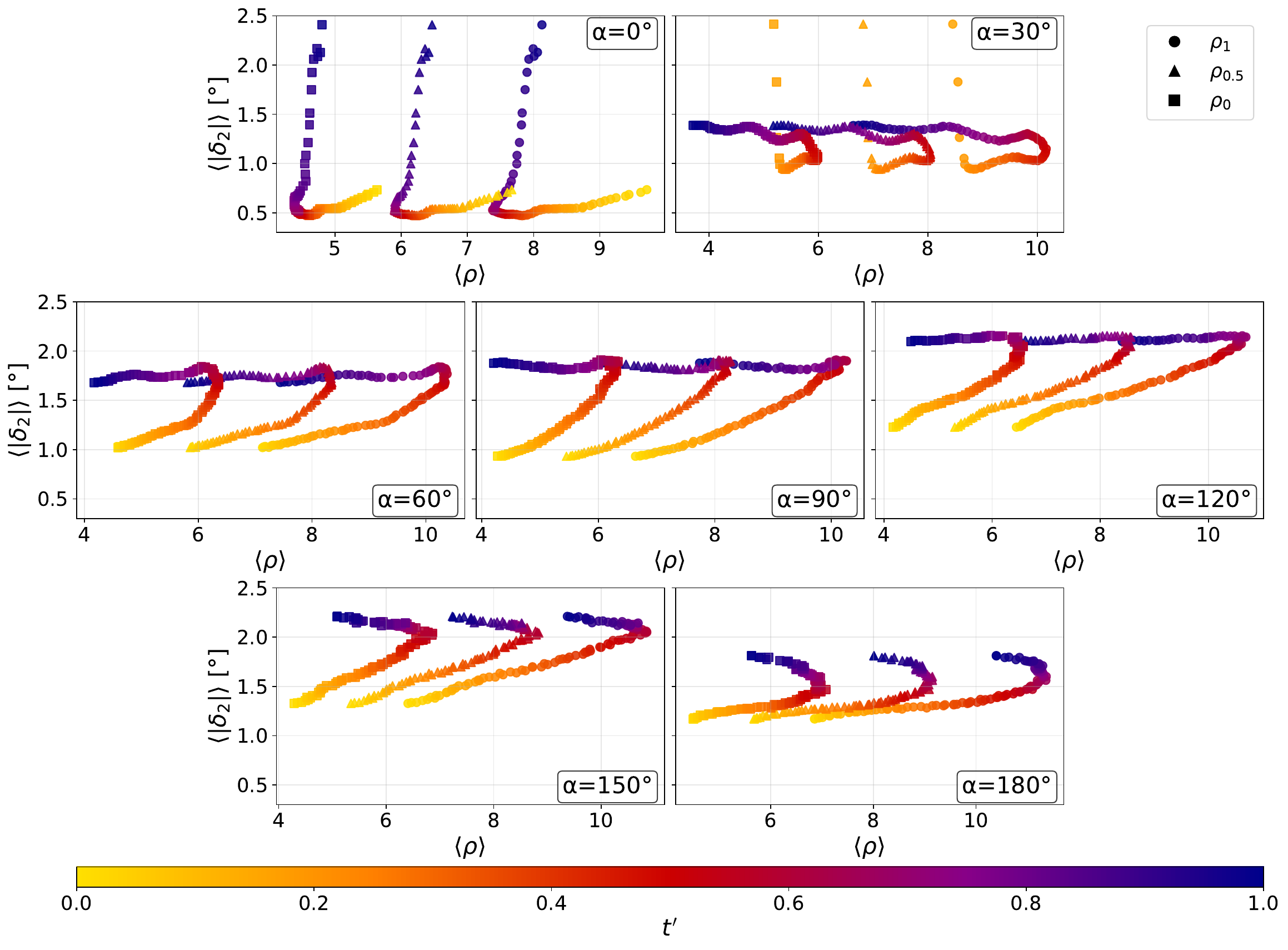}
    \caption{Temporal evolution of mean $|\delta_2|$ and crowdedness for different crossing angles $\alpha$. Each point corresponds to the average deviation $\langle |\delta_2|\rangle$ and crowdedness $\langle \rho \rangle$ of pedestrians at a given scaled time $t^\prime$ during a crossing-flow interaction. The color gradient represents the temporal evolution of the deviation-crowdedness relationship averaged across all trials.}
    \label{fig4}
\end{figure}

\noindent We also investigate how directional adjustments evolve with the local interaction state. We consider the joint temporal evolution of crowdedness and the two deviation measures, $\langle|\delta_1|\rangle$ and $\langle|\delta_2|\rangle$. The resulting phase-space trajectories are shown in Figures \ref{fig3} and \ref{fig4}.\\

\noindent For most crossing angles, the trajectories form open loops rather than collapsing onto a single-valued relationship between crowdedness and directional deviation. The system initially moves toward larger crowdedness as the two groups increasingly overlap, while the deviations evolve along an angle-dependent pathway. At later times, crowdedness decreases as the groups separate, but the trajectories do not retrace their initial paths. Consequently, the same level of crowdedness may be associated with different magnitudes of directional deviation depending on whether the pedestrians are entering the interaction, experiencing the most strongly overlapping configuration, or emerging from it. The loop structures therefore reveal a history dependence of the deviation-crowdedness relationship: crowdedness alone does not uniquely determine the directional state of the flow. As in the velocity-deviation phase space, this non-retracing behavior suggests that the development and dissolution of the crossing interaction are dynamically distinct processes rather than simple time reversals of one another.\\

\noindent The loop-like structures may also be related to the formation and subsequent dissolution of self-organized striped patterns. As crowdedness increases, pedestrians progressively adjust their trajectories while the two streams interact and organize into stripe-like structures. Once such collective organization has developed, pedestrians may maintain their altered directions even as crowdedness begins to decrease. The directional configuration of the crowd therefore retains a memory of the preceding interaction, producing a different return pathway in phase space. The loops should consequently be interpreted as hysteresis-like rather than as conventional hysteresis, since crowdedness is not externally cycled as a control parameter but evolves intrinsically during the transient crossing process.\\

\noindent The two deviation measures exhibit broadly similar loop structures, but with an important difference. The trajectories of $\langle|\delta_1|\rangle$ show a stronger dependence on crossing angle and generally span a larger range of angular deviations. This is consistent with $\delta_1$ measuring the accumulated departure from the expected group direction: as the interaction develops, pedestrians may progressively adopt walking directions that differ substantially from their original group motion. By contrast, the trajectories of $\langle|\delta_2|\rangle$ occupy a considerably narrower angular range and, for several crossing angles, approach comparatively flat late-time branches. Thus, even when pedestrians have developed a substantial deviation from their expected group direction, the changes between successive walking directions remain relatively small. The phase-space representation therefore reinforces the distinction between the two measures: $\delta_1$ reflects the directional displacement accumulated through the interaction, whereas $\delta_2$ captures the smaller, incremental turning adjustments through which this displacement develops.\\

\noindent An additional distinction between the two deviation measures emerges from the direction in which the phase-space loops are traversed. For $\langle\delta_1\rangle$, most trajectories show a clear loop: as $t'$ increases, crowdedness first increases while $\langle\delta_1\rangle$ also rises, but during the later separation phase the trajectory returns through a different branch. This means that, for the same crowdedness, pedestrians may have larger or smaller deviations depending on whether the system is in the build-up or relaxation stage of the crossing interaction. For $\langle\delta_1\rangle$, however, the loops are less pronounced and often become almost flattened. The trajectories mainly move horizontally with crowdedness, while $\langle\delta_1\rangle$ changes only weakly. This suggests that the accumulated deviation from the group direction, $\delta_1$, retains a stronger memory of the interaction history, whereas the incremental turning measure, $\delta2$, responds more locally to the instantaneous interaction state, and remains comparatively more stable.\\

\noindent The dependence on the anisotropy parameter $c$ is again predominantly quantitative rather than qualitative. For each crossing angle, the trajectories obtained using $\rho_1$, $\rho_{0.5}$, and $\rho_0$ retain essentially the same overall shape and temporal progression, but are displaced along the crowdedness axis. Decreasing $c$ reduces the estimated crowdedness by progressively discounting pedestrians outside the field of view, while leaving the underlying deviation dynamics unchanged. The repeated appearance of nearly equivalent phase-space trajectories across the three perception models is therefore consistent with the results reported earlier: perceptual anisotropy primarily rescales the crowdedness measure rather than altering the qualitative organization of the crossing dynamics.\\

\end{document}